\documentclass[aps,prl,twocolumn,superscriptaddress]{revtex4}

\newcommand{\w}{{\omega}}
\newcommand{\s}{{\sigma}}

\def\be{\begin{eqnarray}}
\def\ee{\end{eqnarray}}
\newcommand{\nn}{\nonumber\\}

\newcommand{\Eq}[1]{Eq.~(\ref{#1})}

\newcommand{\ra}{\rightarrow}

\usepackage{graphicx}

\begin{document}

\title{Phase Transition in a Two-level-cavity System in an Ohmic Environment}
\author{Chang-Qin Wu}
\affiliation{Department of Physics, University of California at
Berkeley, Berkeley, CA 94720, USA} \affiliation{Department of
Physics, Fudan University, Shanghai 200433, China}
\author{Jian-Xin Li}
\affiliation{Department of Physics, University of California at
Berkeley, Berkeley, CA 94720, USA} \affiliation{National Laboratory
of Solid State Microstructure, Nanjing University, Nanjing 210093,
China}
\author{Dung-Hai Lee}
\affiliation{Department of Physics, University of California at
Berkeley, Berkeley, CA 94720, USA} \affiliation{Material Science
Division, Lawrence Berkeley National Laboratory, Berkeley, CA 94720,
USA}

\date{\today}

\begin{abstract}
We propose that in the presence of an Ohmic, de-phasing type
environment, a two-level-cavity system undergoes a quantum phase
transition from a state with damped Rabi oscillation to a state
without. We present the phase diagram and make predictions for
pump and probe experiment. Such a strong coupling effect of the
environment is beyond the reach of conventional perturbative
treatment.
\end{abstract}

\maketitle

A combined two-level system and an cavity electromagnetic mode has
been proposed as an important realization of qubit in quantum
information science\cite{raimond,mabuchi,stievater,wall}. In
quantum optics the coherent oscillation, or Rabi oscillation, of
such a combined system is the key to phenomena such as lasing and
electromagnetically induced transparency\cite{Boller,Fleis}. In
both cases the environment acts to destroy the desired quantum
coherence~\cite{leggett,dykman}. As a result, understanding the
coupling to the environment and controlling it has become one of
the most pressing task in quantum control.

In conventional wisdom, the environment broadens quantum levels
and damps the Rabi oscillation. In fact, in the literature, the
life time of the Rabi oscillation is often taken as a measure of
the coupling strength to the environment. In this paper we point
out a non-perturbative effect of the environment which can not be
encapsulated in the level broadening picture.

This effect results in a quantum phase transition as a function of
the ratio of two coupling strength, $\gamma/\lambda$, when the
cavity mode is in resonance with the two-level system. The
numerator, $\gamma$, is the interaction strength between the
two-level system and the cavity mode. The denominator, $\lambda$,
is same quantity between the two-level system and the environment.
The phase at large $\gamma/\lambda$ exhibits damped Rabi
oscillation, while that at small $\gamma/\lambda$ does not. One
can tune through this phase transition by changing the degree of
excitation of the cavity mode. This can be achieved by, e.g.,
varying the Rabi driving strength. The universality class of this
phase transition is that of the inverse-square Ising model in one
dimension. It is triggered by the proliferation of up
$\leftrightarrow$ down flips of the two-level-cavity system. The
purpose of this paper is to make predictions for the manifestation
of this phase transition in an
absorption experiment.\\

{\bf The model~~} In the absence of the environment we use the
Jaynes-Cummings model\cite{cm} to describe the coupling between a
two-level system and a single cavity electromagnetic mode,
\begin{equation}
H_0=t_0\s_z+g(\s_+a+\s_-a^\dagger)+\w_0a^\dagger a.\label{h0}
\end{equation}
Here $\sigma_{z}=\pm 1$ denotes the upper/lower states of the
two-level system, and $a^+(a)$ is the creation (annihilation)
operator of the frequency $\w_0$ cavity photon. In \Eq{h0} $\s_\pm
=\s_x\pm i\s_y$, and $2t_0$ is the energy gap of the  two level
system.

The Hilbert space upon which \Eq{h0} acts is the direct product of
the Hilbert spaces of the two-level system and the cavity photons.
A state in this Hilbert space is labelled by $|s,n\rangle$ where
$s=\pm 1$ denotes the states of the two-level system and $n$
counts the number of cavity photon. It's easy to see that the
matrix corresponds to $H_0$ decouples into independent $2\times 2$
blocks. Each block is spanned by the states $|+,n-1\rangle$ and
$|-,n\rangle$. If we use a pseudospin variable $\tau_z$ to label
these two states so that $|\tau_z=+1\rangle=|+1,n-1\rangle$ and
$|\tau_z=-1\rangle=|-1,n\rangle$, \Eq{h0} becomes
\begin{equation}
H_0=\Delta_0\tau_z+\gamma \tau_x,
\end{equation}
where $\Delta_0=t_0-\w_0/2$. In obtaining the above equation we
have dropped a constant energy term $(n-1/2)\w_0$ and defined
$\gamma\equiv 2g\sqrt{n}$. The quantity $\gamma$ measures the
coupling between the two-level system and the cavity photon. It
can be varied by changing $g$ or $n$. Experimentally reaching
large $\gamma$ by making $n$ large can be achieved by coupling the
two-level system to a strong (hence classical) driving field - the
Rabi driving field.

The environment is modeled by a bath of harmonic oscillators in
the spirit of Caldeira and Leggett~\cite{CL}. In this paper we
consider a specific type of environment whose coupling with the
two-level-cavity system is given by
\begin{equation}
H=\Delta_0\tau_{z}+\gamma
\tau_{x}+\sum_{\alpha}g_{\alpha}\tau_z(b_{\alpha}+b^{\dagger}_{\alpha})
+\sum_{\alpha}\omega_{\alpha}b^{\dagger}_{\alpha}b_{\alpha}.
\label{h}
\end{equation}
Here $\alpha$ labels the environment oscillators and
$\omega_\alpha$ is their frequency. As shown by Caldeira and
Leggett~\cite{CL} the information about the environment is
summarized in the following spectral function
$J(\omega)=\sum_{\alpha}g_{\alpha}^{2}\delta
(\omega-\omega_{\alpha}).$ In the rest of the paper we shall
consider an ``Ohmic'' environment i.e., \be
J(\omega)=\lambda~\omega~ e^{-\omega/\omega_c},\label{ohmic}\ee
where $\lambda$ is a dimensionless coupling constant and
$\omega_c$ is a cut-off frequency.

In perturbative picture the type of coupling in \Eq{h} will give
rise to dephasing. In the jargon of nuclear magnetic resonance the
dephasing lift time $T_2$ measures the strength $\lambda$ in
\Eq{ohmic}. The model in \Eq{h} can be realized by, e.g., a Rabi
driven ``Cooper pair box'' in tunneling contact with a metal. The
type of environmental effect described by \Eq{h} can be simulated by
the charge noise caused by quantum tunnelling of electrons between a
trapping center in the tunnel junction and the metal.\cite{sousa}.
\\

{\bf Perturbative treatment~~} Given that at $t=0$ the two-level
system is in, say, the upper state, the Rabi oscillation will be
manifested as the quantum oscillation of the two-level-cavity system
between $\tau_z=\pm 1$. It can be seen by considering the following
quantity \be P(t)=|\langle +1|e^{-iHt}|+1\rangle|^{2}-|\langle
-1|e^{-iHt}|+1\rangle |^2.\label{p} \ee For simplicity let us
consider the case of resonant driving, i.e., $\omega_0=2t_0$. In the
absence of the coupling to the environment ($g_\alpha=0$) it is
simple to show that $P(t)=\cos(2\gamma t$), the Rabi frequency is
$\pi/\gamma$.
\begin{figure}
\includegraphics[angle=0,scale=0.7]{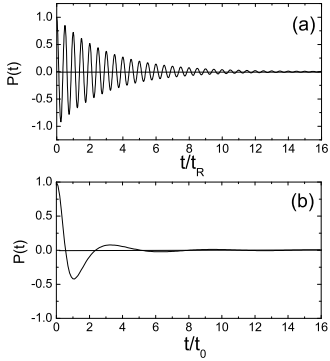}
\caption{(a) The Rabi oscillation $P(t)$ as a function of
$t/(\pi/\gamma)$ calculated using perturbation theory. In producing
this graph we took
$\gamma=0.05t_0,\lambda=0.01,\omega_0=2t_0,\omega_{c}=2t_0$. (b)
$P(t)$ as a function of $t\Delta$ ($\Delta$ is defined in \Eq{mt})
in the instanton metal phase. In producing this figure we took the
$\Omega=\Delta$ in \Eq{mt}.}
\end{figure}
 To the
lowest order, the retarded self-energy due to the coupling to the
environment is given by \be\Sigma_{s's}(\omega)=\langle s'|V
{[\omega-\tilde H_{0}+i\epsilon]}^{-1}V|s\rangle\ee where $s,s'=\pm
1$, $V=
\sum_{\alpha}g_{\alpha}\tau_{z}(b_{\alpha}+b^{\dagger}_{\alpha})$,
and $\tilde H_0=\gamma
\tau_x+\sum_{\alpha}\omega_{\alpha}b^{\dagger}_{\alpha}b_{\alpha}$.
Standard calculation gives \be\Sigma(\w)={1\over
2}\pmatrix{f_1(\w)&f_2(\w)\cr f_2(\w)&f_1(\w)}\ee

where \be &&f_{1}(\w)=\sum_{s}\left[\Lambda(\omega+s\gamma)-i\pi
J(\omega+s\gamma)\theta(\omega+s\gamma)\right]\nn&&f_{2}(\w)=\sum_{s}s\left[\Lambda(\omega+s\gamma)-i\pi
J(\omega+s\gamma)\theta(\omega+s\gamma)\right].\ee  In the above
equations $\theta(\w)$ is the step function, and
\be\Lambda(\omega)=-\lambda\left[\omega_c+\omega
e^{-\omega/\omega_c}\int_{-\w/\w_c}^{\infty}dt~{e^{-t}\over
t}\right].\ee Using the above result and the relation \be\langle
s'|e^{-iHt}|s\rangle=\int_{-\infty}^{\infty} {d\omega\over 2\pi
i}\langle s'\Big|{e^{-i\omega t}\over \omega+i\epsilon -\tilde
H_0-\Sigma(\omega)}\Big|s\rangle, \ee we have calculated $P(t)$
for a typical set of parameters, and the result is shown in
Fig.1(a). Indeed the Rabi
oscillation is damped.\\

{\bf Non-perturbative treatment~~} We can integrate out the
environmental oscillators to obtain the following quantum
partition function for the two-level system~\cite{wll}
\begin{equation}
Z=\sum_{n=0}^\infty \int_0^\beta d\tau_{2n} \int_0^{\tau_{2n}}
d\tau_{2n-1}\cdots \int_0^{\tau_{2}} d\tau_{1}
~\gamma^{2n}~e^{-S[Q(\tau_i)]} \label{z}
\end{equation}
Here $2n$ counts the number of $\tau_z\ra -\tau_z$ flips of the
two-level-cavity system occurring between imaginary time $\tau=0$
and $\tau=\beta$. The quantity $Q(\tau_i)$ is an alternating
sequence of $\pm 1$ ($+1$ for down $\ra$ up flip and $-1$ for up
$\ra$ down flip) at $\tau_1,..,\tau_{2n}$. Setting $\hbar\ra 1$
the action $S[Q(\tau_i)]$ in \Eq{z} is given by \be S=\sum_{i\ne
j}Q_i G(\tau_{ij})Q_j+\Delta_0\sum_i
(Q_i-1)(\tau_{i+1}-\tau_i),\label{action}\ee where \be
G(\tau)=\int_0^\infty d\omega
{J(\omega)\over\w^2}\exp(-\omega|\tau|)=\lambda\ln
\left(|\tau|/\tau_0\right)\ee with $\tau_0=1/\w_c$. In the limit
of $t\tau_0<<1$ and $\Delta_0\tau_0<<1$  perturbative
renormalization group calculation gives, \be
&&\frac{d\lambda}{dl}=-2\lambda
\gamma^{2},~~\frac{d\gamma}{dl}=\left(1-{\lambda\over
2}\right)\gamma\nn&&{d\Delta_0\over dl}=\Delta_0(1-\gamma^2).\ee
here $l=\ln(\tau/\tau_0)$ is the running cutoff scale. For
$\w_0=2t_0$ these equations predict a phase transition between an
instanton (i.e.,$\tau_z$ flipping event) insulator phase and an
instanton metal phase. In the former phase $\gamma$ is
renormalized to zero at large time scale, while in the latter it
diverges.

Alternatively one can map Eqs.(\ref{z},\ref{action}) onto the
classical partition function of the inverse square ferromagnetic
Ising model in one dimension \be \beta H=-K\sum_{i,j} {a^2\over
|x_i-x_j|^2}\s_i\s_j-h\sum_i\s_i,\label{ising}\ee where $a$ is the
lattice constant. By computing the Boltzmann weight associated with
domain wall configurations it can be shown that the lattice constant
$a$ in the Ising model plays the role of $\tau_0$ and \be
\gamma=e^{-c_1 K},~~~~\lambda=c_2K,~~{\rm and}~~ \Delta_0={2h/
a},\label{corr}\ee where $c_1$ and $c_2$ are order unity constants.
The advantage of mapping onto the inverse square Ising model is that
for $h=0$ extensive Monte Carlo simulation has been performed on
such model~\cite{lui}. This will allow us to access the physics in
the $t\tau_0\sim 1$ regime.

Now let us briefly review the simulation results of \Eq{ising} at
$h=0$. As a function of $K$ there are two phases. The weak $K$ phase
is a paramagnet and the strong $K$ phase is a ferromagnet. Due to
the long-range nature of the spin-spin interaction the phase
transition occurs at finite $K$ despite of one-dimensionality. In
the paramagnetic phase the spin-spin correlation decays
exponentially, $\langle\s_i\s_j\rangle\sim \left({a\over
|x_i-x_j|}\right)^2 e^{-|x_i-x_j|/\xi}$, and the correlation length
diverges as $\xi/a\sim e^{b/\sqrt{K_c-K}},$ like in the
Kosterlitz-Thouless transition. The ferromagnetic phase has non-zero
$\langle\s_i\rangle$. At the ferromagnetic $\ra$ paramagnetic
transition $\langle\s_i\rangle$ undergoes a discontinuous jump to
zero. The value of $\langle\s_i\rangle$ at the critical point is
universal and is equal to $1/2$. In the ferromagnetic phase the
connected part of the spin-spin correlation function decays
algebraically with distance, i.e.,
$\Gamma(x_i-x_j)=\langle\s_i\s_j\rangle-\langle\s_i\rangle\langle\s_j\rangle\sim{a\over
|x_i-x_j|^\eta}.$ The decay exponent $\eta$ is expected to be equal
to $4\sqrt{(K-K_c)/K_c}$. At the critical point $\Gamma(x)\sim
1/\ln{|{x\over a}|}$. In the presence of a non-zero magnetic field,
one expects $\langle\s_i\rangle \ne 0$ for all $K$ values. The phase
transition will not be present because the global $\s_i\ra -\s_i$
symmetry is explicitly broken. In this case one expects
$\Gamma(x)\sim \left({a\over |x|}\right)^\eta e^{-|x|/\xi}$.\\

{\bf Predictions for optical absorption experiment} In this section
we use the results discussed in the last section to predict the
outcome of optical absorption experiment. (Note that if strong
$\gamma$ is achieved by Rabi driving, this would amount to a pump
and probe experiment). Imagine coupling the two-level system to the
following AC probe \be H_{probe}=g_c\cos(\w t)\tau_z.\ee By applying
the Fermi golden rule, it is simple to show that the absorption
spectrum is given by \be A(\w)=g_c^2\sum_\alpha
|\langle\alpha|\tau_z|0\rangle|^2\delta(\w-E_\alpha+E_0).\ee In the
above $|0\rangle$ is the ground state and $|\alpha\rangle$ are the
exact eigenstates of the combined two-level-environment system. It
can be shown simply that the zero-temperature imaginary time
spin-spin correlation function is the Laplace transform of $A(\w)$:
$\langle\tau_z(\tau)\tau_z(0)\rangle=\int_0^\infty d\w e^{-\w
\tau}A(\w)/g_c^2.$ Equivalently the absorption spectrum is the
inverse Laplace transform of $\langle\tau_z(\tau)\tau_z(0)\rangle$.
Thus the knowledge of the behavior of the spin-spin correlation
function in the last section enable us to make explicit predictions
about the absorption spectrum.

First we consider the case of resonance, i.e., $\w_0=2t_0$. This
corresponds to the abscissa,$\gamma/\lambda$, of Fig.~2.  According
to \Eq{corr} this is equal to $e^{-c_1 K}/c_2 K$ of the Ising model.
Thus large/small $\gamma/\lambda$ corresponds to a small/large $K$.
As a result we expect the system to be in the instanton metal phase
for large $\gamma/\lambda$ and in the instanton insulator phase in
the opposite limit. After inverse Laplace transforming the spin-spin
correlation function we obtain the following prediction for the
optical absorption spectrum at low frequencies
\begin{equation}
A(\omega)=k_1\theta(\omega-\Delta)(\omega-\Delta)e^{-\w/\Omega}\label{mt}
\end{equation}
in the instanton metal phase, and
\begin{equation}
A(\omega)=\Phi\left({\gamma/\lambda}\right)\delta(\omega)+k_2\omega^{\eta-1}
e^{-\w/\Omega}. \label{it}
\end{equation}
in the instanton insulator phase. In Eqs.(\ref{mt}) and (\ref{it}),
$k_{1,2}$ are constants, $\Omega$ is the absorption bandwidth,
$\Phi(\gamma/\lambda)=\langle\s_i\rangle$ is the magnetization of
the Ising model. \Eq{mt} predicts the presence of an absorption gap
$\Delta$ in the instanton metal phase. This gap is related to the
correlation length in the Ising model by $\Delta \sim
{1\over\tau_0}{a\over\xi}.$  As the Ising correlation length
diverges at the critical point $\Delta$ closes according to
\be\Delta\sim \tau_0^{-1}
\exp[-b/\sqrt{(\gamma/\lambda)-(\gamma/\lambda)_c}].\ee At
criticality a zero-frequency delta function in the absorption
spectrum abruptly develops and the finite frequency absorption has a
$1/\w$ dependence. Throughout the instanton insulator phase, \Eq{it}
predicts the presence of a zero-frequency delta function peak and a
gapless absorption spectrum. 

The case of off-resonance ($\Delta_0\ne 0$) corresponds to the
presence of a magnetic field in the Ising model. Translating the
behavior of Ising  spin-spin correlation into $A(\w)$, we obtain
\begin{equation}
A(\omega)=\Phi\left({\gamma/
\lambda},\Delta_0\right)\delta(\omega)+k_{3}
\theta(\omega-\Delta)(\omega-\Delta)^{\eta-1}e^{-\omega/\Omega}
\label{off}
\end{equation}
Similar to the instanton insulator phase, a zero-frequency delta
function is present. On the other hand, the non-zero frequency
absorption shows a gap similar to the instanton metal phase. By
fixing $\gamma/\lambda$ at a low value and tuning $\w_0$ one will
see the closing of the absorption gap when $\w_0$ reaches the
resonance value.

One might think that reaching the critical point on the abscissa of
Fig.~2 requires the fine tuning of two parameters. This is in fact
not true, because the entire $\Delta_0=0$,  $0<\gamma/\lambda
<(\gamma/\lambda)_c$ interval corresponds to a line of fixed points.
This is similar to the low temperature critical line of the
Kosterlitz-Thouless phase transition. To reach this critical line
one simply sets $\gamma/\lambda$ at a low value and adjusts the
detuning parameter $\Delta_0$. In terms of optical absorption, the
critical line is reached when the finite frequency absorption gap
closes. This enables one to tune to resonance (i.e., the abscissa of
Fig.~2) easily. Once this is achieved, one can tune across
$(\gamma/\lambda)_c$ by adjust the Rabi driving strength.

\begin{figure}
\includegraphics[angle=0,scale=0.6]{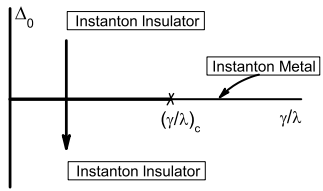}
\caption{A schematic phase diagram. The thick line in the abscissa
is the line of critical points discussed in the test. All regions
except the thin line in the abscissa are in the instanton
insulator phase. A scan along the line with an arrow will exhibit
the closing of the absorption gap when cross the
abscissa.}\label{pd}
\end{figure}

To study the Rabi oscillation in the instanton metal phase we need
to compute the $P(t)$ in \Eq{p} in the presence of an environment.
To do it non-perturbatively  we connect it to the correlation
functions of the Ising model. Let $|0\rangle$ be the ground state
of \Eq{h}. The initial observation of $\tau_z=+1$ projects the
state of the system to $\left(1+\tau_z\over 2\right)|0\rangle.$
Using the time evolution operator associated with \Eq{h} to evolve
this state and computing the $\tau_z$ expectation value at a later
time $t$ we obtain \be P(t)&&={1\over 2 \langle
0|1+\tau_z|0\rangle}\{\langle 0|\tau_{z}(t)|0\rangle +\langle
0|\tau_{z}\tau_{z}(t)\tau_{z}|0\rangle \nn&&+\langle
0|\tau_{z}\tau_{z}(t)|g\rangle+ \langle
0|\tau_{z}(t)\tau_{z}|g\rangle\},\ee where
$\tau_z(t)=e^{iHt}\tau_z e^{-iHt}$. It is important to note that
in the above equation $t$ is the real (not imaginary) time. At
$\Delta_0=0$ and in the absence of spontaneous symmetry breaking
$\langle 0|\tau_{z}(t)|0\rangle=\langle
0|\tau_{z}\tau_{z}(t)\tau_z|0\rangle=0$. As a result
\begin{equation}
P(t)=\langle 0|\tau_{z}\tau_{z}(t)|g\rangle+ \langle
0|\tau_{z}(t)\tau_{z}|g\rangle=\int_{0}^{\infty}\cos(\omega
t)A(\omega)d\omega.\label{ps}
\end{equation}
Inserting \Eq{mt} into \Eq{ps}, we  calculated the Rabi
oscillation in the instanton metal phase, the result is presented
in Fig.1(b). One can see a damped Rabi oscillations qualitative
similar to the perturbative result shown in Fig.1(a). In the
instanton insulator phase, the strong coupling with the
environment locks the the two-level-cavity system, and the Rabi
oscillations is complectly quenched.

In summary, we have shown that although the environment is harmful
to quantum coherence, it contains interesting physics in its own
right. In particular, we have demonstrated that  when the
environment is Ohmic and only causes dephasing, there is a quantum
phase transition between a state which exhibits damped Rabi
oscillation to a state which does not. In reality the
dephasing-only requirement of our model has been achieved
partially already in experiment. For example, for the ``Cooper
pair box'' a $T_1$ (life time due to the up $\leftrightarrow$ down
flipping of the two-level system) which is an order of magnitude
longer than $T_2$ (the dephasing life time) has been
achieved~\cite{siddiqi}. In the langauge of Ising model,
$T_1/\tau_0$ imposes a finite size cutoff to the critical behavior
discussed earlier. Hopefully future technological progresses will
enable one to probe the regime where $T_1>>T_2$, hence the novel
1D analog of the Kosterlitz-Thouless critical behavior.

\acknowledgements

DHL thanks R. de Sousa and I. Siddiqi for helpful discussions.  JXL
and CQW thank the support of the Berkeley Scholar program and the
NSF of China. JXL and CQW were also supported by MSTC(2006CBOL1002)
and MOE of China (B06011), respectively. DHL was supported by the
Directior, Office of Science, Office of Basic Energy Sciences,
Materials Sciences and Engineering Division, of the U.S. Department
of Energy under Contract No. DE-AC02-05CH11231.

\end{document}